\documentclass[aps,prl,twocolumn,superscriptaddress,nofootinbib]{revtex4-1}

\usepackage{xcolor}
\definecolor{tab20green}{rgb}{0.1725490196,0.6274509804,0.1725490196}
\definecolor{tab20red}{rgb}{0.8392156863,0.1529411765,0.1568627451}
\definecolor{tab20orange}{rgb}{1.0,0.4980392157,0.05490196078}
\definecolor{tab20blue}{rgb}{0.1215686275,0.4666666667,0.7058823529}
\usepackage{amsfonts,amssymb,amsmath} 
\usepackage{graphicx}
\usepackage{dsfont}
\usepackage{soul}
\usepackage{ulem}
\usepackage[
  colorlinks,
  linkcolor=blue,
  urlcolor=blue,
  citecolor=blue,
  plainpages=false,
  pdfpagelabels,
]{hyperref}

\graphicspath{{./figs/}}

\DeclareMathAlphabet\mathbfcal{OMS}{cmsy}{b}{n}

\newcommand{\baleqn}{\begin{equation}\begin{aligned}[b]}
\newcommand{\ealeqn}{\end{aligned}\end{equation}}
\newcommand{\baleqns}{\begin{equation*}\begin{aligned}}
\newcommand{\ealeqns}{\end{aligned}\end{equation*}}
\newcommand{\be}{\begin{equation}}
\newcommand{\ee}{\end{equation}}

\newcommand{\beq}{\begin{equation}}
\newcommand{\eeq}{\end{equation}}
\newcommand{\beqa}{\begin{eqnarray}}
\newcommand{\eeqa}{\end{eqnarray}}

\begin{document}
\title{Seeing topological edge and bulk currents in time-of-flight images}

\author{Alvaro Rubio-Garc\'{i}a\hyperlink{thankslink1}{\textsuperscript{\textdagger}}\hypertarget{thankslink2}{}}
\affiliation{Instituto de Estructura de la Materia IEM-CSIC, Calle Serrano 123, Madrid E-28006, Spain}
\author{Chris N. Self\hyperlink{thankslink1}{\textsuperscript{\textdagger}}\hypertarget{thankslink2}{}}
\affiliation{School of Physics and Astronomy, University of Leeds, Leeds LS2 9JT, UK}
\author{Juan Jose Garc\'{i}a-Ripoll}
\affiliation{Instituto de F\'{i}sica Fundamental IFF-CSIC, Calle Serrano 113b, Madrid E-28006, Spain}
\author{Jiannis K. Pachos}
\affiliation{School of Physics and Astronomy, University of Leeds, Leeds LS2 9JT, UK}

\date{\today}
\pacs{...}

\begin{abstract}
Here we provide a general methodology to directly measure the topological currents emerging in the optical lattice implementation of the Haldane model. Alongside the edge currents supported by gapless edge states, transverse currents can emerge in the bulk of the system whenever the local potential is varied in space, even if it does not cause a phase transition. In optical lattice implementations the overall harmonic potential that traps the atoms provides the boundaries of the topological phase that supports the edge currents, as well as providing the potential gradient across the topological phase that gives rise to the bulk current. Both the edge and bulk currents are resilient to several experimental parameters such as trapping potential, temperature and disorder. We propose to investigate the properties of these currents directly from time-of-flight images with both short-time and long-time expansions.

\end{abstract}

\maketitle
{\let\thefootnote\relax\footnote{{\hypertarget{thankslink1}{}\hyperlink{thankslink2}{\textsuperscript{\textdagger}} AR-G and CNS contributed equally to this work.}}}

{\bf Introduction:--} 
Two-dimensional topological insulators are described by a non-zero integer topological index, typically given by the Chern number, $\nu$~\cite{kane2005quantum,kane2005z2topological,bernevig2006quantum,hasan2010colloquium,qi2011topological}. At physical boundaries of the system the Chern number changes from a non-zero value inside the material to zero value outside it. This change can be interpreted as a topological phase transition, which is manifested at the boundary of the topological system as a one-dimensional gapless edge state~\cite{kane2005quantum,kane2005z2topological,bernevig2006quantum,hasan2010colloquium,qi2011topological}. As the change in the Chern number necessarily takes only integer values rather than the continuum the edge states are thus robust to small perturbations and finite temperature. In the presence of a constant chemical potential $V(\boldsymbol{r})=V$ the particle-hole imbalance in the population of the edge states gives rise to the edge current
\begin{equation}
\label{eq:current1}
I_\text{edge} = {\nu V \over 2\pi},
\end{equation}
traversing along the boundary of the system~\cite{hatsugai1993chern,hao2008topological,colomes2018antichiral,self2019topological}. The edge currents provide a powerful tool to experimentally probe the topological properties of the system. 

Alongside the edge currents, the bulk of the topological insulator can support currents, whenever the local potential has a non-zero gradient~\cite{self2019topological,lensky2015topological,geller1994currents}. In contrast to the edge currents, these bulk currents can appear even in the absence of gapless modes, with the system remaining gapped at all times. In our previous paper~\cite{self2019topological} we showed that the bulk of a topological insulator on the lattice is formed by hidden many-body entangled edge states. The individual currents of these states cancel each other, when the bulk is homogeneous, giving zero flux. However, we showed that an inhomogeneous potential in the system can disentangle these edge states and form a net current perpendicular to the local gradient of that potential, with an intensity given by
\begin{equation}
\label{eq:current2}
I_\text{bulk} = \frac{\nu}{2\pi}\,a_0|\boldsymbol{\nabla} V(\boldsymbol{r})|,
\end{equation}
with $a_0$ the lattice constant. Similar to the edge currents the bulk currents are robust against temperature and local disorder. Unlike the edge currents the position and direction of the topological bulk currents can be controlled at will by external potentials, thus offering a unique platform for developing new technologies.

A physical system where edge and bulk currents can naturally emerge is optical lattice implementations of the Haldane model~\cite{tarruell2012creating,jotzu2014experimental,goldman2016topological,flaschner2016experimental,asteria2019measuring}. Time-of-flight images enable the measurement of the Chern number of this system thus allowing us to confirm its topological nature~\cite{alba2011seeing}. In these experiments harmonic potentials are used to keep the atoms localised in a certain region, in addition to the Hamiltonian of the Haldane model. This trapping causes a spatial variation in the local potential. Here we demonstrate that, together with the edge currents at the boundary of the topological phase, bulk currents also emerge due to the variation of the trapping potential across the topological phase of the system. The interplay between edge and bulk currents can be obtained by time-of-flight (TOF) images making their investigation directly accessible to current experiments. We find signatures of the edge and bulk currents in both the short-time and long-time expansion images, thus offering novel and versatile means of probing topological current physics.

\begin{figure}[t]
	\centering
	\includegraphics[width=0.9\columnwidth]{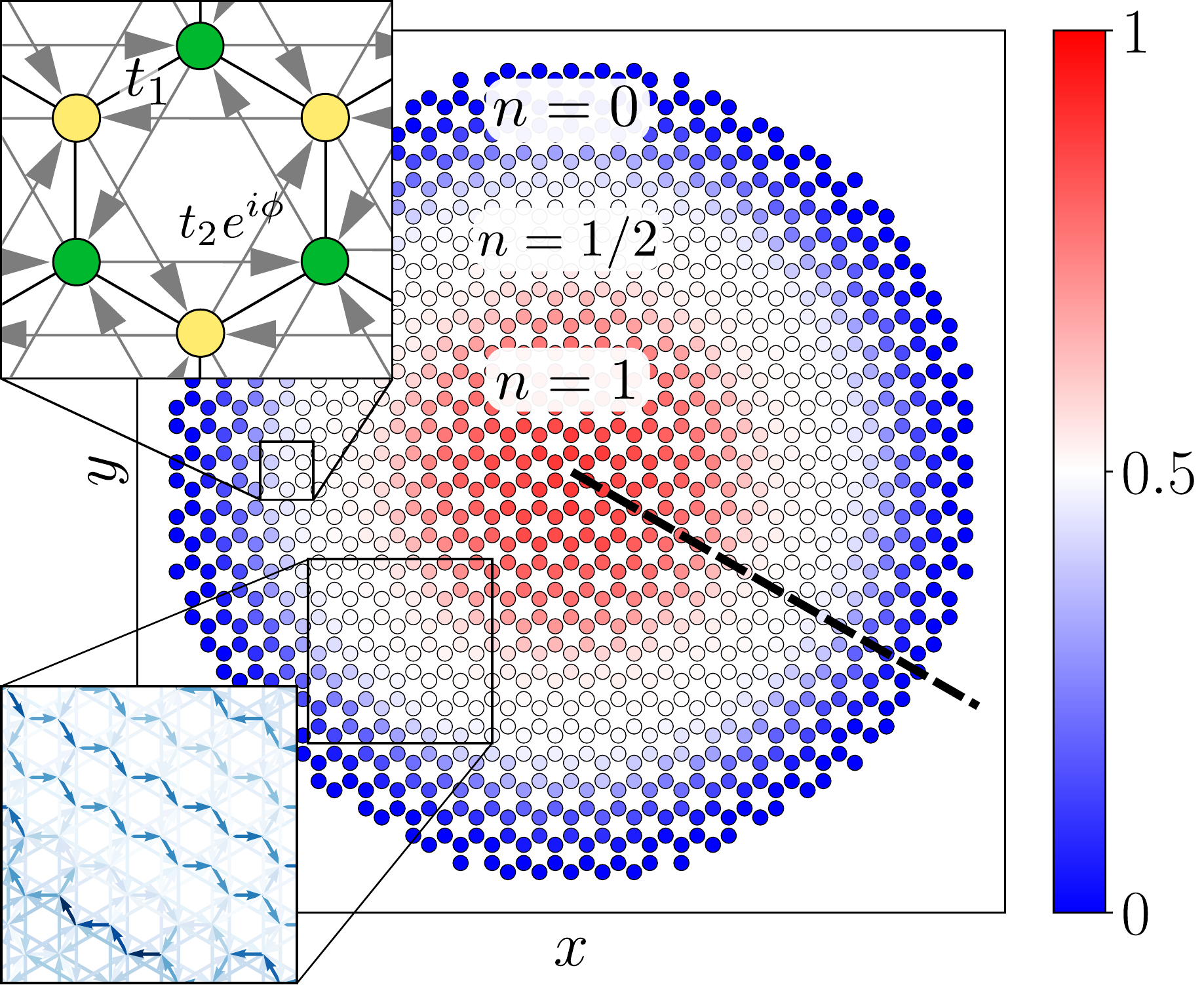}
	\caption[current]{Real space occupations of the Haldane model inside a harmonic trap. There are three clearly visible phases with densities $n=1$ (red), $n=1/2$ (white), and $n = 0$ (blue). We are interested in the flux of the currents crossing the radial dashed line. (Upper detail) Hexagonal plaquette of the Haldane lattice model. Arrows denote hopping directions for which $\nu_{ij} = +1$ in Eq.~(\ref{eqn:haldane-model-ham}). (Lower detail) Real space currents (arrows) between lattice sites. Intensity of the arrow colour denotes the intensity of the current. Parallel edge currents are visible at the boundary of the topological phase, while the current in the bulk has the opposite orientation. Data shown is for $T=0$ and a lattice with $1,200$ sites. The harmonic potential (\ref{harmonic}) has $V_0 = -2.1$, $k=5.6$ and $r_\text{max} = 13\sqrt{3}$.}
	\label{fig:real_space_densities}
\end{figure}

{\bf Realising topological edge and bulk currents in optical lattices:--} To investigate the properties of the topological edge and bulk currents we consider the implementation of the Haldane model with ultra-cold atoms in optical lattices. The Haldane model is defined on a hexagonal lattice with a fermionic mode $\lbrace c_i,\ c^\dagger_i\rbrace$ living at each vertex of the lattice. The Hamiltonian is
\begin{align}
H &= \sum_{\langle ij\rangle} t_1 \, c^\dagger_i c_j + \sum_{\langle\langle ij\rangle\rangle} t_2 \, \textrm{e}^{i\nu_{ij}\phi} c^\dagger_i c_j + \sum_i V(\boldsymbol{r}_i) \, c^\dagger_i c_i,
\label{eqn:haldane-model-ham}
\end{align}
where $t_1$ and $t_2$ are the nearest and next-nearest neighbour hopping strengths, respectively, $\phi$ is a complex phase, $\nu_{ij}$ a parameter that is $\pm 1$ depending on the direction of the hopping $i\rightarrow j$, as shown in Fig.~\ref{fig:real_space_densities}, and $V(\boldsymbol{r})$ is a local potential. For certain choices of parameters the Haldane model has a topological phase with Chern number $\nu = \pm 1$ as well as a trivial phase where $\nu =0 $ \cite{haldane1988model}. All results here are produced for the parameters $t_1 = 1$, $t_2 = 0.1$, $\phi = \pi/2$ that, at half-filling and for $V(\boldsymbol{r})=0$ give $\nu=1$. 

\begin{figure}[t]
\centering

\includegraphics[width=\columnwidth]{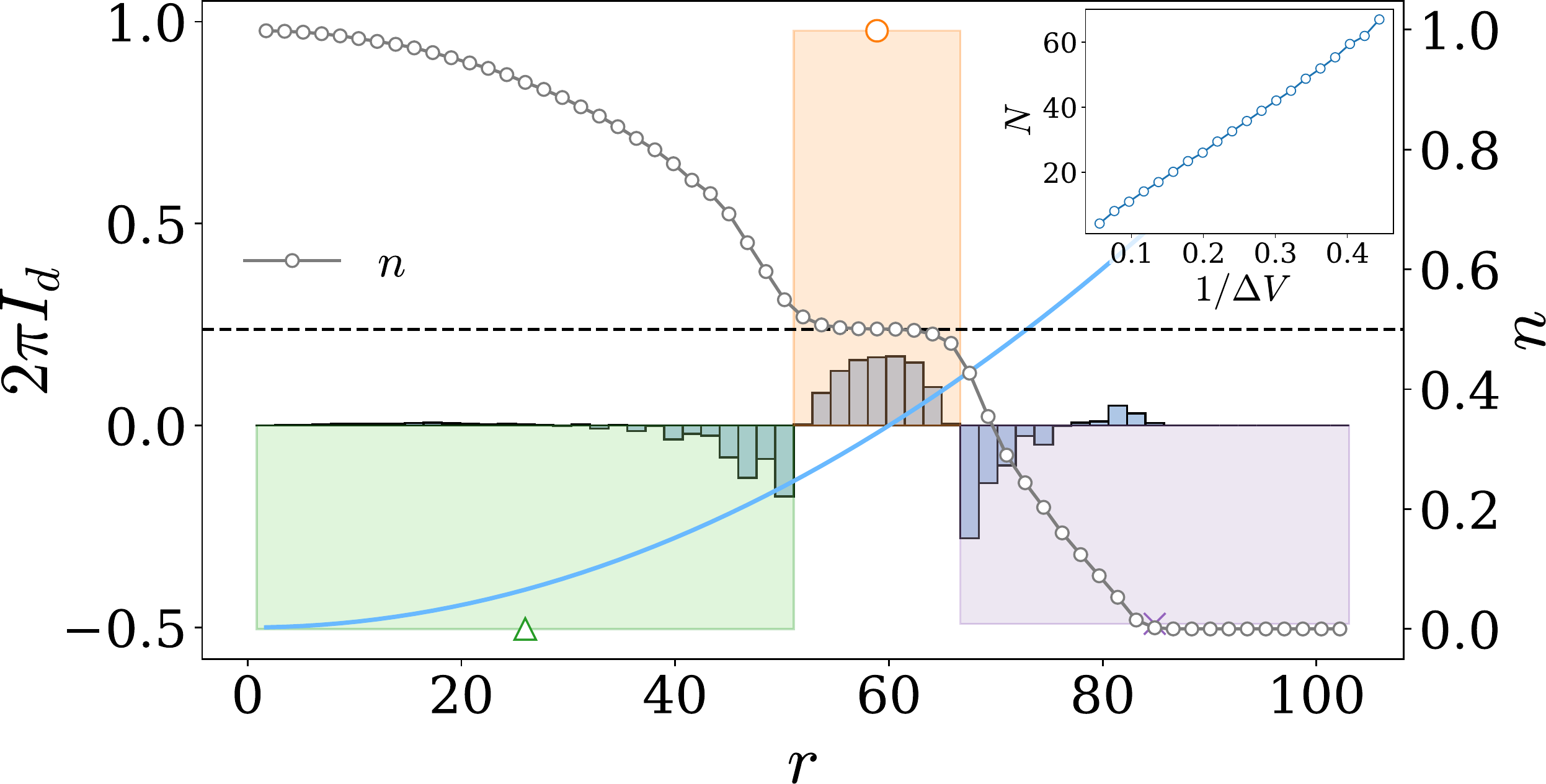}
\caption[current]{Occupation of lattice sites (circles) along a radial line -- as shown in Fig.~\ref{fig:real_space_densities} -- inside an harmonic potential trap (blue line), and local currents across small bins (bars) in the same radial line, with $r$ the distance from the centre of the trap. Three phases are clearly visible from the occupations: $n=1$ at the centre of the trap, $n=1/2$ in the middle corresponding to an annulus of the Haldane phase and $n=0$ at the far end of the system. The total edge and bulk currents resulting from the sum of the small bins are indicated with the three large bars. The total edge currents are equal in magnitude, propagating in the same direction (negative value), while the bulk current through the $n=1/2$ region is opposite (positive value) with double magnitude. Data presented is for $T=0$ and a lattice size of roughly $28,800$ sites. The harmonic potential (\ref{harmonic}) has $V_0 = -3$, $k= 9$ and $r_\textrm{max} = 60\sqrt{3}$. (Inset) The width $N$ of the topological phase increases linearly with $1/\Delta V$, where $\Delta V$ is the potential difference at the two boundaries of the phase.}
\label{fig:optical-lattices}
\end{figure}

Motivated by initial theoretical work~\cite{alba2011seeing} several experimental realisations of the Haldane model in optical lattices have been achieved~\cite{tarruell2012creating,jotzu2014experimental,goldman2016topological,flaschner2016experimental,asteria2019measuring}. In these experiments the Haldane lattice model given by (\ref{eqn:haldane-model-ham}) is engineered inside a harmonic potential trap that keeps the atoms localised in a certain region. We describe the harmonic trap as 
\begin{equation}
\label{harmonic}
V(r) = V_0 + k\left({r\over r_\text{max}}\right)^2,
\end{equation} 
where $V_0$ is an overall chemical potential, while $k$ and $r_\text{max}$ are suitable parameters of the trap.
Depending on the choice of the constants $V_0$ and $k$ this potential can give rise to a `wedding cake' structure, with three phases emerging at different radii from the centre of the harmonic trap, as shown in Fig.~\ref{fig:real_space_densities}. Near the centre, $r=0$, a trivial phase is nested with lattice filling fraction given by $n=1$ and Chern number $\nu=0$ as all bands of the system are populated. Next an annulus emerges with half-filling, $n=1/2$, that corresponds to the Haldane phase with Chern number $\nu=\pm1$. Finally, far from the centre, where the harmonic potential is too large, a trivial $\nu=0$ phase is present due to zero band population, $n=0$. The topologically non-trivial annulus configuration has two boundaries neighbouring the trivial $n=1$ and $n=0$ phases. Moreover, the trapping potential varies radially across the topological phase, as seen in Fig.~\ref{fig:optical-lattices}. Hence, due to (\ref{eq:current1}) and (\ref{eq:current2}) we expect edge and bulk currents to naturally emerge without additional engineering of the system. 

The density current, $J_{ij}$, flowing between two sites $i$ and $j$ of the lattice are dictated by the continuity equation for single site occupation~\cite{self2019topological}. From that we can determine the distribution of the current flowing across a cross section of the lattice that goes radially from the centre of the trap, as shown in Fig.~\ref{fig:real_space_densities}. The local profile of currents supported in the system is shown as a histogram in Fig.~\ref{fig:optical-lattices}. We observe that as the harmonic potential has a smooth profile the edge states expand over a large range of the system. Moreover, the smooth change of the local potential in the bulk causes the bulk currents to disperse over the whole topological phase. 

In Fig.~\ref{fig:optical-lattices} we also plot the total edge and bulk currents as the sum of the local ones;  it is clear that the two edge currents are equal and flowing in the same direction while the bulk current is twice their value and flowing in the opposite direction. This behaviour is expected from assuming that the topological phase is stable in the middle of the bulk, where the local potential favouring half filling $n=1/2$, while the potential increases as we move further out and decreases as we move closer to the centre of the trap. In this case two opposite potentials are formed at the inner and outer boundaries of the system causing the parallel edge currents due to (\ref{eq:current1}). At the same time the linear interpolation of the potential between the two boundaries causes a bulk current to be formed as dictated by (\ref{eq:current2}). Finally, we observe in Fig.~\ref{fig:optical-lattices} (Inset) that the width of the Haldane phase increases inversely proportional to the potential difference, $\Delta V$, between the two boundaries. This possible control in the macroscopic properties of the system can facilitate the manipulation of the density and spatial extension of the currents under investigation in order to improve their visibility.

\begin{figure}[t]
	\centering
	\includegraphics[width=\columnwidth]{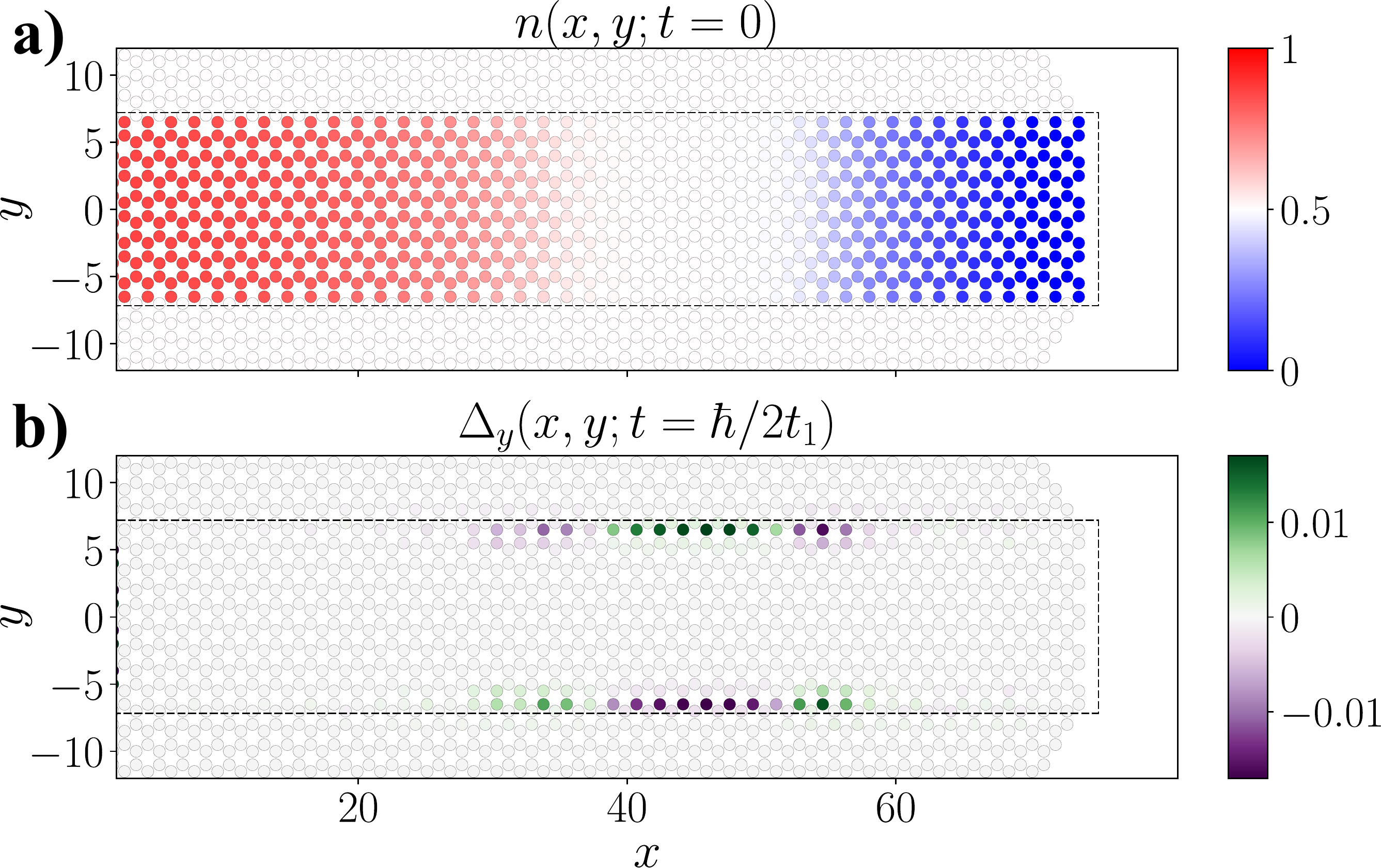}
	\caption[tof_differences]{(a) Densities of a sample of sites (inside dashed line) at a stationary state inside an harmonic trap isolated by a mask. Densities outside the sample are taken to be 0. (b) Density difference $\Delta_y (x, y; t) = n(x, y; t) - n(x, -y; t)$ after a short-time expansion ($t = \hbar/2t_1$) of the sites within the sample when the trap is turned off. A clear pattern of density imbalances is already observed, that corresponds to one up-going current inside the Haldane phase and two down-going currents at its boundaries. Data shown is for $T=0$ and a lattice with $9800$ sites. The employed harmonic potential (\ref{harmonic}) has $V_0 = -2.1$, $k=5.8$ and $r_\text{max} = 43\sqrt{3}$.}
	\label{fig:ste_density_differences}
\end{figure} 

{\bf Detecting density currents with time-of-flight images:--} The density currents of the atoms in the $n=1/2$ phase of the optical lattice, as well as at its boundaries, can be directly measured with TOF images. To distinguish between edge and bulk currents we want to selectively obtain information about particular regions of the optical lattice. We can achieve that with TOF measurements performed by switching off the Hamiltonian terms within particular regions of the system. As a result, the atoms in these regions freely expand and can be detected. This allows information, such as the atom density and velocity distribution, to be obtained. Releasing a small sample of lattice sites around the Haldane phase can therefore tell us about the physics of the edge and bulk currents. 

We would like to investigate both the short-time and long-time TOF images. A short-time expansion of a sample of sites can already reveal the existence of the bulk and edge currents. In this case we take a horizontal sample of sites, as shown in Fig.~\ref{fig:ste_density_differences} (a), when the system is in a stationary state inside a harmonic trap. Then, we remove the trap and let the atoms inside that sample freely expand. After sufficiently small times a distinctive population pattern is formed, as shown in Fig.~\ref{fig:ste_density_differences} (b). We observe that there are more particles at the top of the sample in the bulk of the Haldane phase region, while at its boundaries the particles accumulate at the bottom of the sample. These particle density differences point to the existence of a particle density current propagating perpendicular to the local gradient of the trap inside the topological phase in accordance with (\ref{eq:current2}) and two counter propagating currents at the topological phase boundaries as dictated by (\ref{eq:current1}).

\begin{figure}[t]
	\centering
	\includegraphics[width=\columnwidth]{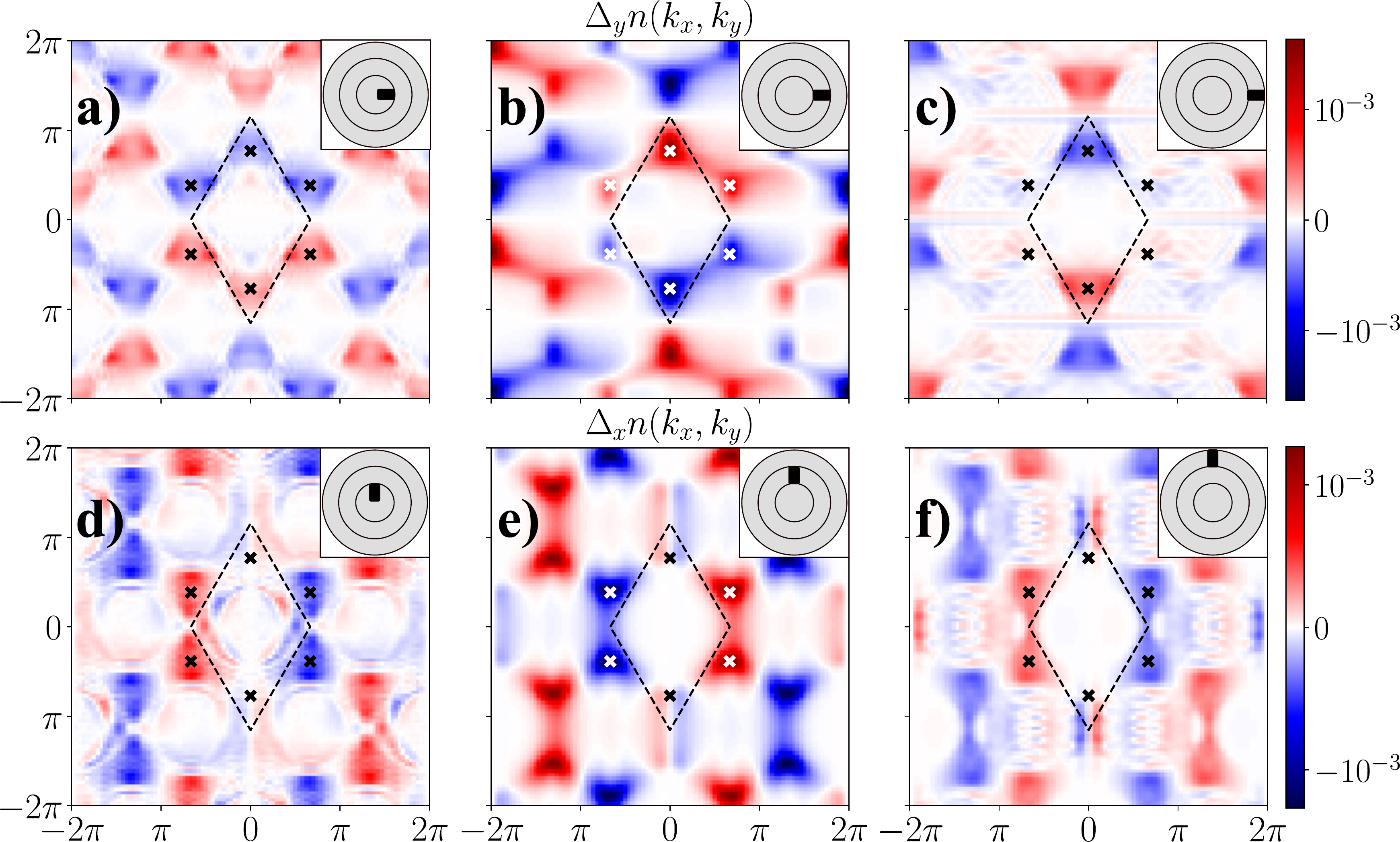}
	\caption[tof_differences]{Density differences in the TOF images taken at different phases of the system. The dotted lines mark the Brillouin zone of the hexagonal lattice and the crosses show the location of the Dirac points. (Insets) Sketch of the rectangular region where particles were sampled for the TOF image. The circular coronas show the three topological phases in the system. (a-c) Difference $\Delta_y n(k_x,k_y) = n(k_x,k_y)-n(k_x,-k_y)$ in the $y$-direction of the momentum for samples taken along the $x$-axis of the lattice. Clearly there are currents moving along the vertical direction, with the bulk currents (b) traveling opposite to the edge ones (a, c). (d-f) Difference $\Delta_x n(k_x,k_y) = n(k_x,k_y)-n(-k_x,k_y)$ in the $x$-direction of the momentum for samples taken along the $y$ axis of the lattice. There are currents moving along the horizontal direction, with the bulk currents traveling opposite to the edge ones. Data shown is for $T=0$ and a lattice with $9800$ sites. The employed harmonic potential (\ref{harmonic}) has $V_0 = -2.0$, $k=5.5$ and $r_\text{max} = 38\sqrt{3}$.}
	\label{fig:tof_density_differences}
\end{figure}

To obtain the signature of density currents from the long-time TOF images we select smaller regions of the lattice that isolate the bulk from the edge of the topological phase. Differences in the velocity distribution over the Brillouin zone for different regions of the trapped sample can signal a net flux of particles traveling in a certain direction. This is shown in Fig.~\ref{fig:tof_density_differences} (a-c), where the colormap pictures the difference $\Delta_y (k_x, k_y) = n(k_x, k_y) - n(k_x, -k_y)$ in the TOF sampling of rectangular patches along the $x$-axis of the optical lattice [see Inset]. A clear difference in the momentum population with opposite $k_y$ is seen, indicating the existence of currents moving along the $y$-direction of the lattice. Moreover, the direction of movement in (b) is opposite to that of (a, c), in agreement with Fig.~\ref{fig:optical-lattices}. Figs. \ref{fig:tof_density_differences} (d-f) show the density difference $\Delta_x (k_x, k_y) = n(k_x, k_y) - n(-k_x, k_y)$ for samples along the $y$-axis of the lattice. In this case, currents flow along the $x$-axis, with the direction of the bulk currents (e) opposite to the direction of the edge ones (d, f).

\begin{figure}[t]
\centering
\includegraphics[width=\columnwidth]{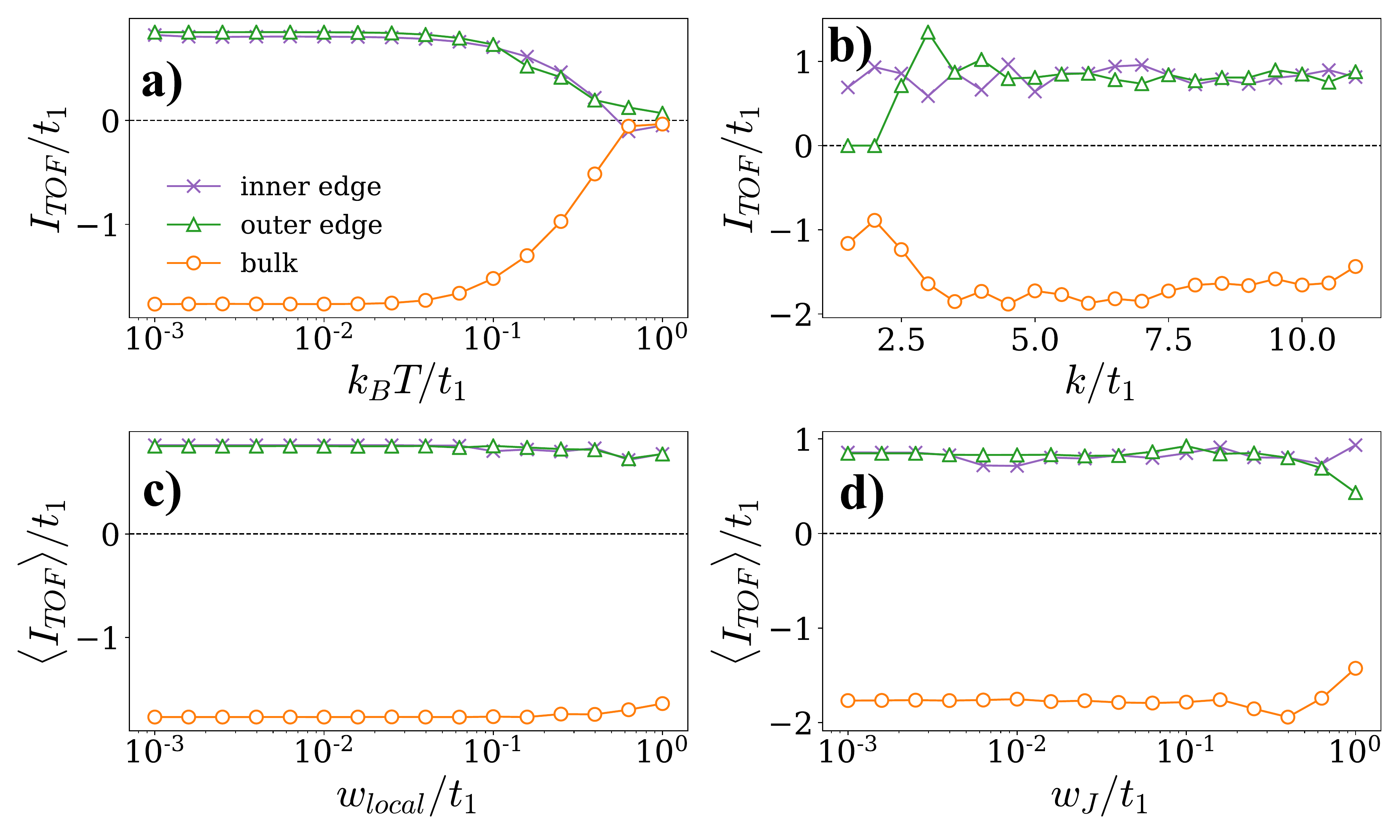}
\caption[current]{(a) Stability of the edge and bulk TOF currents against the temperature of the optical lattice. Currents decay when the temperature reaches the energy gap of the Haldane model. (b) TOF currents when the trap strength $k/t_1$ is widened. For $k/t_1 > 2.5$ all three phases are present in the system and the edge and bulk currents are stable. (c, d) Mean TOF currents over 15 samples with local (c) and coupling (d) disorder in the trap. The mean currents remain stable even when the scale of the disorder is comparable to the energy gap. Data shown is for $T=0$ and a lattice with $9800$ sites. The harmonic potential \ref{harmonic} has $V_0 = -2.0$, $k=5.5$ and $r_\text{max} = 38\sqrt{3}$.}
\label{fig:optical-resilience}
\end{figure}

To determine the behaviour of the currents observed in TOF images under realistic conditions met in experiments we study the stability of the currents against temperature, local potential and coupling disorder as well as variations in the strength of the harmonic potential. These effects are commonly present in optical lattice implementations. As the edge and bulk currents have a topological origin we expect them to be both largely resilient against these perturbations~\cite{self2019topological}. To proceed we define the net TOF current as
\begin{equation}
\boldsymbol{I}_\text{TOF} = \sum_{\boldsymbol{k}\in\textrm{BZ}} \boldsymbol{k}\, n\left(\boldsymbol{k}\right)
\label{eqn:tof_currents}
\end{equation}
where we weight all the momenta inside the Brillouin zone with the density distribution $n\left(\boldsymbol{k}\right)$ measured by the TOF images. We sample regions of the lattice as in Fig.~\ref{fig:tof_density_differences} to distinguish between bulk and edge density currents. Fig.~\ref{fig:optical-resilience}(a) shows the behavior of the TOF currents (\ref{eqn:tof_currents}) against the temperature of the system. We observe that the topological currents remain intact until the temperature is comparable to the energy gap of the Haldane model. In Fig.~\ref{fig:optical-resilience}(b) we test the stability of the currents against the strength of the trap potential (\ref{harmonic}) parametrised by $k/t_1$. For wide enough traps capable of supporting the three phases of the system, we observe stable TOF currents with intensity independent on the trap potential. In Fig.~\ref{fig:optical-resilience}(c) we add a local disordered potential in the Hamiltonian with values randomly sampled from the uniform distribution $[-w_{local}/2t_1, w_{local}/2t_1]$, and in Fig.~\ref{fig:optical-resilience}(d) we multiply each coupling of the model with a value drawn from the uniform distribution $[1-w_J/2t_1, 1+w_J/2t_1]$ modelling imperfections in designed Hamiltonian. In both cases we observe that the mean edge and bulk currents are resilient to all these forms of disorder.

{\bf Conclusions:--} To summarise, we have presented a general and versatile way to investigate the currents of a topological insulator with TOF images. In particular, we focused on the behaviour of the Haldane model simulated by ultra-cold atoms in optical lattices. In the presence of inhomogeneous potentials arising e.g. by the harmonic trapping of the atoms topological bulk currents emerge along the edge currents. The bulk currents are proportional to the gradient of the local potential that naturally arises in the optical lattice experiment due to the harmonic trapping. The topological origin of both the edge and bulk currents make them robust against perturbations, such as  inhomogeneous potentials superimposed on top of the lattice, errors in the exact values of the couplings of the Haldane model, variations in the trapping potential of the atomic cloud as well as finite temperature. As bulk currents do not need phase transitions to be generated they are more versatile than the edge currents and easier to engineer and manipulate. Hence, their implementation with optical lattices would open the way to employ them for quantum technologies.

\begin{acknowledgments}
We would like to thank Monika Aidelsburger and Sofyan Iblisdir for inspiring conversations. This work was supported by the EPSRC grant EP/R020612/1, Spanish Projects PGC2018-094792-B-I00 (MCIU/AEI/FEDER, EU), PGC2018-094180-B-I00 (MCIU/AEI/FEDER, EU), FIS2015-63770-P (MINECO/FEDER, EU), CAM/FEDER Project No.
S2018/TCS-4342 (QUITEMAD-CM) and CSIC Research Platform PTI-001. Statement of compliance with EPSRC policy framework on research data: This publication is theoretical work that does not require supporting research data.

\end{acknowledgments}

\bibliography{references}

\begin{thebibliography}{18}%
\makeatletter
\providecommand \@ifxundefined [1]{%
 \@ifx{#1\undefined}
}%
\providecommand \@ifnum [1]{%
 \ifnum #1\expandafter \@firstoftwo
 \else \expandafter \@secondoftwo
 \fi
}%
\providecommand \@ifx [1]{%
 \ifx #1\expandafter \@firstoftwo
 \else \expandafter \@secondoftwo
 \fi
}%
\providecommand \natexlab [1]{#1}%
\providecommand \enquote  [1]{``#1''}%
\providecommand \bibnamefont  [1]{#1}%
\providecommand \bibfnamefont [1]{#1}%
\providecommand \citenamefont [1]{#1}%
\providecommand \href@noop [0]{\@secondoftwo}%
\providecommand \href [0]{\begingroup \@sanitize@url \@href}%
\providecommand \@href[1]{\@@startlink{#1}\@@href}%
\providecommand \@@href[1]{\endgroup#1\@@endlink}%
\providecommand \@sanitize@url [0]{\catcode `\\12\catcode `\$12\catcode
  `\&12\catcode `\#12\catcode `\^12\catcode `\_12\catcode `\%12\relax}%
\providecommand \@@startlink[1]{}%
\providecommand \@@endlink[0]{}%
\providecommand \url  [0]{\begingroup\@sanitize@url \@url }%
\providecommand \@url [1]{\endgroup\@href {#1}{\urlprefix }}%
\providecommand \urlprefix  [0]{URL }%
\providecommand \Eprint [0]{\href }%
\providecommand \doibase [0]{http://dx.doi.org/}%
\providecommand \selectlanguage [0]{\@gobble}%
\providecommand \bibinfo  [0]{\@secondoftwo}%
\providecommand \bibfield  [0]{\@secondoftwo}%
\providecommand \translation [1]{[#1]}%
\providecommand \BibitemOpen [0]{}%
\providecommand \bibitemStop [0]{}%
\providecommand \bibitemNoStop [0]{.\EOS\space}%
\providecommand \EOS [0]{\spacefactor3000\relax}%
\providecommand \BibitemShut  [1]{\csname bibitem#1\endcsname}%
\let\auto@bib@innerbib\@empty
\bibitem [{\citenamefont {Kane}\ and\ \citenamefont
  {Mele}(2005{\natexlab{a}})}]{kane2005quantum}%
  \BibitemOpen
  \bibfield  {author} {\bibinfo {author} {\bibfnamefont {C.~L.}\ \bibnamefont
  {Kane}}\ and\ \bibinfo {author} {\bibfnamefont {E.~J.}\ \bibnamefont
  {Mele}},\ }\href {\doibase 10.1103/PhysRevLett.95.226801} {\bibfield
  {journal} {\bibinfo  {journal} {Phys. Rev. Lett.}\ }\textbf {\bibinfo
  {volume} {95}},\ \bibinfo {pages} {226801} (\bibinfo {year}
  {2005}{\natexlab{a}})}\BibitemShut {NoStop}%
\bibitem [{\citenamefont {Kane}\ and\ \citenamefont
  {Mele}(2005{\natexlab{b}})}]{kane2005z2topological}%
  \BibitemOpen
  \bibfield  {author} {\bibinfo {author} {\bibfnamefont {C.~L.}\ \bibnamefont
  {Kane}}\ and\ \bibinfo {author} {\bibfnamefont {E.~J.}\ \bibnamefont
  {Mele}},\ }\href {\doibase 10.1103/PhysRevLett.95.146802} {\bibfield
  {journal} {\bibinfo  {journal} {Phys. Rev. Lett.}\ }\textbf {\bibinfo
  {volume} {95}},\ \bibinfo {pages} {146802} (\bibinfo {year}
  {2005}{\natexlab{b}})}\BibitemShut {NoStop}%
\bibitem [{\citenamefont {Bernevig}\ and\ \citenamefont
  {Zhang}(2006)}]{bernevig2006quantum}%
  \BibitemOpen
  \bibfield  {author} {\bibinfo {author} {\bibfnamefont {B.~A.}\ \bibnamefont
  {Bernevig}}\ and\ \bibinfo {author} {\bibfnamefont {S.-C.}\ \bibnamefont
  {Zhang}},\ }\href {\doibase 10.1103/PhysRevLett.96.106802} {\bibfield
  {journal} {\bibinfo  {journal} {Phys. Rev. Lett.}\ }\textbf {\bibinfo
  {volume} {96}},\ \bibinfo {pages} {106802} (\bibinfo {year}
  {2006})}\BibitemShut {NoStop}%
\bibitem [{\citenamefont {Hasan}\ and\ \citenamefont
  {Kane}(2010)}]{hasan2010colloquium}%
  \BibitemOpen
  \bibfield  {author} {\bibinfo {author} {\bibfnamefont {M.~Z.}\ \bibnamefont
  {Hasan}}\ and\ \bibinfo {author} {\bibfnamefont {C.~L.}\ \bibnamefont
  {Kane}},\ }\href {\doibase 10.1103/RevModPhys.82.3045} {\bibfield  {journal}
  {\bibinfo  {journal} {Rev. Mod. Phys.}\ }\textbf {\bibinfo {volume} {82}},\
  \bibinfo {pages} {3045} (\bibinfo {year} {2010})}\BibitemShut {NoStop}%
\bibitem [{\citenamefont {Qi}\ and\ \citenamefont
  {Zhang}(2011)}]{qi2011topological}%
  \BibitemOpen
  \bibfield  {author} {\bibinfo {author} {\bibfnamefont {X.-L.}\ \bibnamefont
  {Qi}}\ and\ \bibinfo {author} {\bibfnamefont {S.-C.}\ \bibnamefont {Zhang}},\
  }\href {\doibase 10.1103/RevModPhys.83.1057} {\bibfield  {journal} {\bibinfo
  {journal} {Rev. Mod. Phys.}\ }\textbf {\bibinfo {volume} {83}},\ \bibinfo
  {pages} {1057} (\bibinfo {year} {2011})}\BibitemShut {NoStop}%
\bibitem [{\citenamefont {Hatsugai}(1993)}]{hatsugai1993chern}%
  \BibitemOpen
  \bibfield  {author} {\bibinfo {author} {\bibfnamefont {Y.}~\bibnamefont
  {Hatsugai}},\ }\href@noop {} {\bibfield  {journal} {\bibinfo  {journal}
  {Physical review letters}\ }\textbf {\bibinfo {volume} {71}},\ \bibinfo
  {pages} {3697} (\bibinfo {year} {1993})}\BibitemShut {NoStop}%
\bibitem [{\citenamefont {Hao}\ \emph {et~al.}(2008)\citenamefont {Hao},
  \citenamefont {Zhang}, \citenamefont {Wang}, \citenamefont {Zhang},\ and\
  \citenamefont {Wang}}]{hao2008topological}%
  \BibitemOpen
  \bibfield  {author} {\bibinfo {author} {\bibfnamefont {N.}~\bibnamefont
  {Hao}}, \bibinfo {author} {\bibfnamefont {P.}~\bibnamefont {Zhang}}, \bibinfo
  {author} {\bibfnamefont {Z.}~\bibnamefont {Wang}}, \bibinfo {author}
  {\bibfnamefont {W.}~\bibnamefont {Zhang}}, \ and\ \bibinfo {author}
  {\bibfnamefont {Y.}~\bibnamefont {Wang}},\ }\href@noop {} {\bibfield
  {journal} {\bibinfo  {journal} {Physical Review B}\ }\textbf {\bibinfo
  {volume} {78}},\ \bibinfo {pages} {075438} (\bibinfo {year}
  {2008})}\BibitemShut {NoStop}%
\bibitem [{\citenamefont {Colom\'es}\ and\ \citenamefont
  {Franz}(2018)}]{colomes2018antichiral}%
  \BibitemOpen
  \bibfield  {author} {\bibinfo {author} {\bibfnamefont {E.}~\bibnamefont
  {Colom\'es}}\ and\ \bibinfo {author} {\bibfnamefont {M.}~\bibnamefont
  {Franz}},\ }\href {\doibase 10.1103/PhysRevLett.120.086603} {\bibfield
  {journal} {\bibinfo  {journal} {Phys. Rev. Lett.}\ }\textbf {\bibinfo
  {volume} {120}},\ \bibinfo {pages} {086603} (\bibinfo {year}
  {2018})}\BibitemShut {NoStop}%
\bibitem [{\citenamefont {Self}\ \emph {et~al.}(2019)\citenamefont {Self},
  \citenamefont {Rubio-Garc\'ia}, \citenamefont {Garc\'ia-Ripoll},\ and\
  \citenamefont {Pachos}}]{self2019topological}%
  \BibitemOpen
  \bibfield  {author} {\bibinfo {author} {\bibfnamefont {C.~N.}\ \bibnamefont
  {Self}}, \bibinfo {author} {\bibfnamefont {A.}~\bibnamefont
  {Rubio-Garc\'ia}}, \bibinfo {author} {\bibfnamefont {J.~J.}\ \bibnamefont
  {Garc\'ia-Ripoll}}, \ and\ \bibinfo {author} {\bibfnamefont {J.~K.}\
  \bibnamefont {Pachos}},\ }\href@noop {} {\enquote {\bibinfo {title}
  {Topological currents in the bulk in the absence of gapless states},}\ }
  (\bibinfo {year} {2019}),\ \Eprint {http://arxiv.org/abs/arXiv:1906.01705}
  {arXiv:1906.01705} \BibitemShut {NoStop}%
\bibitem [{\citenamefont {Lensky}\ \emph {et~al.}(2015)\citenamefont {Lensky},
  \citenamefont {Song}, \citenamefont {Samutpraphoot},\ and\ \citenamefont
  {Levitov}}]{lensky2015topological}%
  \BibitemOpen
  \bibfield  {author} {\bibinfo {author} {\bibfnamefont {Y.~D.}\ \bibnamefont
  {Lensky}}, \bibinfo {author} {\bibfnamefont {J.~C.~W.}\ \bibnamefont {Song}},
  \bibinfo {author} {\bibfnamefont {P.}~\bibnamefont {Samutpraphoot}}, \ and\
  \bibinfo {author} {\bibfnamefont {L.~S.}\ \bibnamefont {Levitov}},\ }\href
  {\doibase 10.1103/PhysRevLett.114.256601} {\bibfield  {journal} {\bibinfo
  {journal} {Phys. Rev. Lett.}\ }\textbf {\bibinfo {volume} {114}},\ \bibinfo
  {pages} {256601} (\bibinfo {year} {2015})}\BibitemShut {NoStop}%
\bibitem [{\citenamefont {Geller}\ and\ \citenamefont
  {Vignale}(1994)}]{geller1994currents}%
  \BibitemOpen
  \bibfield  {author} {\bibinfo {author} {\bibfnamefont {M.~R.}\ \bibnamefont
  {Geller}}\ and\ \bibinfo {author} {\bibfnamefont {G.}~\bibnamefont
  {Vignale}},\ }\href {\doibase 10.1103/PhysRevB.50.11714} {\bibfield
  {journal} {\bibinfo  {journal} {Phys. Rev. B}\ }\textbf {\bibinfo {volume}
  {50}},\ \bibinfo {pages} {11714} (\bibinfo {year} {1994})}\BibitemShut
  {NoStop}%
\bibitem [{\citenamefont {Tarruell}\ \emph {et~al.}(2012)\citenamefont
  {Tarruell}, \citenamefont {Greif}, \citenamefont {Uehlinger}, \citenamefont
  {Jotzu},\ and\ \citenamefont {Esslinger}}]{tarruell2012creating}%
  \BibitemOpen
  \bibfield  {author} {\bibinfo {author} {\bibfnamefont {L.}~\bibnamefont
  {Tarruell}}, \bibinfo {author} {\bibfnamefont {D.}~\bibnamefont {Greif}},
  \bibinfo {author} {\bibfnamefont {T.}~\bibnamefont {Uehlinger}}, \bibinfo
  {author} {\bibfnamefont {G.}~\bibnamefont {Jotzu}}, \ and\ \bibinfo {author}
  {\bibfnamefont {T.}~\bibnamefont {Esslinger}},\ }\href
  {https://doi.org/10.1038/nature10871} {\bibfield  {journal} {\bibinfo
  {journal} {Nature}\ }\textbf {\bibinfo {volume} {483}},\ \bibinfo {pages}
  {302 EP } (\bibinfo {year} {2012})}\BibitemShut {NoStop}%
\bibitem [{\citenamefont {Jotzu}\ \emph {et~al.}(2014)\citenamefont {Jotzu},
  \citenamefont {Messer}, \citenamefont {Desbuquois}, \citenamefont {Lebrat},
  \citenamefont {Uehlinger}, \citenamefont {Greif},\ and\ \citenamefont
  {Esslinger}}]{jotzu2014experimental}%
  \BibitemOpen
  \bibfield  {author} {\bibinfo {author} {\bibfnamefont {G.}~\bibnamefont
  {Jotzu}}, \bibinfo {author} {\bibfnamefont {M.}~\bibnamefont {Messer}},
  \bibinfo {author} {\bibfnamefont {R.}~\bibnamefont {Desbuquois}}, \bibinfo
  {author} {\bibfnamefont {M.}~\bibnamefont {Lebrat}}, \bibinfo {author}
  {\bibfnamefont {T.}~\bibnamefont {Uehlinger}}, \bibinfo {author}
  {\bibfnamefont {D.}~\bibnamefont {Greif}}, \ and\ \bibinfo {author}
  {\bibfnamefont {T.}~\bibnamefont {Esslinger}},\ }\href
  {https://doi.org/10.1038/nature13915} {\bibfield  {journal} {\bibinfo
  {journal} {Nature}\ }\textbf {\bibinfo {volume} {515}},\ \bibinfo {pages}
  {237 EP } (\bibinfo {year} {2014})}\BibitemShut {NoStop}%
\bibitem [{\citenamefont {Goldman}\ \emph {et~al.}(2016)\citenamefont
  {Goldman}, \citenamefont {Budich},\ and\ \citenamefont
  {Zoller}}]{goldman2016topological}%
  \BibitemOpen
  \bibfield  {author} {\bibinfo {author} {\bibfnamefont {N.}~\bibnamefont
  {Goldman}}, \bibinfo {author} {\bibfnamefont {J.~C.}\ \bibnamefont {Budich}},
  \ and\ \bibinfo {author} {\bibfnamefont {P.}~\bibnamefont {Zoller}},\ }\href
  {https://doi.org/10.1038/nphys3803} {\bibfield  {journal} {\bibinfo
  {journal} {Nature Physics}\ }\textbf {\bibinfo {volume} {12}},\ \bibinfo
  {pages} {639 EP } (\bibinfo {year} {2016})}\BibitemShut {NoStop}%
\bibitem [{\citenamefont {Fl{\"a}schner}\ \emph {et~al.}(2016)\citenamefont
  {Fl{\"a}schner}, \citenamefont {Rem}, \citenamefont {Tarnowski},
  \citenamefont {Vogel}, \citenamefont {L{\"u}hmann}, \citenamefont
  {Sengstock},\ and\ \citenamefont {Weitenberg}}]{flaschner2016experimental}%
  \BibitemOpen
  \bibfield  {author} {\bibinfo {author} {\bibfnamefont {N.}~\bibnamefont
  {Fl{\"a}schner}}, \bibinfo {author} {\bibfnamefont {B.~S.}\ \bibnamefont
  {Rem}}, \bibinfo {author} {\bibfnamefont {M.}~\bibnamefont {Tarnowski}},
  \bibinfo {author} {\bibfnamefont {D.}~\bibnamefont {Vogel}}, \bibinfo
  {author} {\bibfnamefont {D.-S.}\ \bibnamefont {L{\"u}hmann}}, \bibinfo
  {author} {\bibfnamefont {K.}~\bibnamefont {Sengstock}}, \ and\ \bibinfo
  {author} {\bibfnamefont {C.}~\bibnamefont {Weitenberg}},\ }\href {\doibase
  10.1126/science.aad4568} {\bibfield  {journal} {\bibinfo  {journal}
  {Science}\ }\textbf {\bibinfo {volume} {352}},\ \bibinfo {pages} {1091}
  (\bibinfo {year} {2016})}\BibitemShut {NoStop}%
\bibitem [{\citenamefont {Asteria}\ \emph {et~al.}(2019)\citenamefont
  {Asteria}, \citenamefont {Tran}, \citenamefont {Ozawa}, \citenamefont
  {Tarnowski}, \citenamefont {Rem}, \citenamefont {Fl{\"a}schner},
  \citenamefont {Sengstock}, \citenamefont {Goldman},\ and\ \citenamefont
  {Weitenberg}}]{asteria2019measuring}%
  \BibitemOpen
  \bibfield  {author} {\bibinfo {author} {\bibfnamefont {L.}~\bibnamefont
  {Asteria}}, \bibinfo {author} {\bibfnamefont {D.~T.}\ \bibnamefont {Tran}},
  \bibinfo {author} {\bibfnamefont {T.}~\bibnamefont {Ozawa}}, \bibinfo
  {author} {\bibfnamefont {M.}~\bibnamefont {Tarnowski}}, \bibinfo {author}
  {\bibfnamefont {B.~S.}\ \bibnamefont {Rem}}, \bibinfo {author} {\bibfnamefont
  {N.}~\bibnamefont {Fl{\"a}schner}}, \bibinfo {author} {\bibfnamefont
  {K.}~\bibnamefont {Sengstock}}, \bibinfo {author} {\bibfnamefont
  {N.}~\bibnamefont {Goldman}}, \ and\ \bibinfo {author} {\bibfnamefont
  {C.}~\bibnamefont {Weitenberg}},\ }\href {\doibase 10.1038/s41567-019-0417-8}
  {\bibfield  {journal} {\bibinfo  {journal} {Nature Physics}\ } (\bibinfo
  {year} {2019}),\ 10.1038/s41567-019-0417-8}\BibitemShut {NoStop}%
\bibitem [{\citenamefont {Alba}\ \emph {et~al.}(2011)\citenamefont {Alba},
  \citenamefont {Fernandez-Gonzalvo}, \citenamefont {Mur-Petit}, \citenamefont
  {Pachos},\ and\ \citenamefont {Garcia-Ripoll}}]{alba2011seeing}%
  \BibitemOpen
  \bibfield  {author} {\bibinfo {author} {\bibfnamefont {E.}~\bibnamefont
  {Alba}}, \bibinfo {author} {\bibfnamefont {X.}~\bibnamefont
  {Fernandez-Gonzalvo}}, \bibinfo {author} {\bibfnamefont {J.}~\bibnamefont
  {Mur-Petit}}, \bibinfo {author} {\bibfnamefont {J.~K.}\ \bibnamefont
  {Pachos}}, \ and\ \bibinfo {author} {\bibfnamefont {J.~J.}\ \bibnamefont
  {Garcia-Ripoll}},\ }\href {\doibase 10.1103/PhysRevLett.107.235301}
  {\bibfield  {journal} {\bibinfo  {journal} {Phys. Rev. Lett.}\ }\textbf
  {\bibinfo {volume} {107}},\ \bibinfo {pages} {235301} (\bibinfo {year}
  {2011})}\BibitemShut {NoStop}%
\bibitem [{\citenamefont {Haldane}(1988)}]{haldane1988model}%
  \BibitemOpen
  \bibfield  {author} {\bibinfo {author} {\bibfnamefont {F.~D.~M.}\
  \bibnamefont {Haldane}},\ }\href {\doibase 10.1103/PhysRevLett.61.2015}
  {\bibfield  {journal} {\bibinfo  {journal} {Phys. Rev. Lett.}\ }\textbf
  {\bibinfo {volume} {61}},\ \bibinfo {pages} {2015} (\bibinfo {year}
  {1988})}\BibitemShut {NoStop}%
\end{thebibliography}%




\end{document}